\documentclass{pasj00}
\color{black}
 
\title{Inhomogeneity in the Hot Intracluster Medium of \\
Abell 1060 Observed with Chandra}
\author{Akira~{\sc Hayakawa}\altaffilmark{1}, Tae~{\sc
Furusho}\altaffilmark{2}, Noriko Y.~{\sc Yamasaki}\altaffilmark{2}, 
Manabu~{\sc Ishida}\altaffilmark{1}, and Takaya~{\sc Ohashi}\altaffilmark{1}}
\altaffiltext{1}{Department of Physics, Tokyo Metropolitan University,\\
1-1 Minami-Ohsawa, Hachioji, Tokyo 192-0397}
\altaffiltext{2}{Institute of Space and Astronautical Science, 
Japan Aerospace Exploration Agency, \\ 3-1-1 
Yoshinodai, Sagamihara, Kanagawa 229-8510} 
\email[AH]{akira\_h@phys.metro-u.ac.jp}
\KeyWords{galaxies: clusters: individual (Abell 1060) --- galaxies:
intergalactic medium --- X-rays: galaxies}
\Received{ 2004 June 11} \Accepted{2004 August 3} \Published{}

\begin{document}
\maketitle
\begin{abstract}
A Chandra observation of the non-cooling flow cluster A 1060 has
confirmed that the hot intracluster medium has fairly uniform
distributions of temperature and metal abundance from a radius of
about 230 kpc to the central 5 kpc region ($H_0= 75 {\rm \ km\ s^{-1}\
Mpc^{-1}}$). The radial temperature profile shows a broad peak at
30--40 kpc from the center at a level $\sim20\%$ higher than that in
the outer region. Assuming spatially uniform temperature and abundance
distributions, we derived a 3-dimensional density structure by
iteratively correcting the $\beta$ model, and obtained the central gas
density to be $8.2^{+1.8}_{-1.0} \times 10^{-3}$ cm$^{-3}$. The
distribution of gravitational mass was estimated from the density
profile, and a central concentration of mass within a radius of 50 kpc
was indicated. The data also suggest several high-abundance
regions. The most significant blob adjacent to the central galaxy NGC
3311 has a radius of about 9 kpc, which indicates a metallicity of
$\sim 1.5$ solar.  We consider that this blob may be produced by the
gas stripped off from NGC 3311.
\end{abstract}

\section{Introduction} 

The spatial distributions of the temperature and metallicity in the
intracluster medium (ICM) tell us about the past history of cluster
formation and the metal-enrichment process. In the cores of relaxed
clusters, recent studies from Chandra and XMM-Newton observatories
have revealed various interesting features. Regarding the temperature
structure, the observed absence of very cool gas at the center
strongly requires some heat-injection mechanism (e.g.\
\cite{peterson03}), which is not yet understood very well. A close
study of a system without any significant cooling flow may give new
insight into these problems.  Another important aspect of the
non-cooling-flow system is a reliable estimation of the gravitational
mass profile, based on a fairly uniform temperature distribution. Such
an attempt was carried out for NGC~1399 \citep{ikebe96}, NGC~4636
\citep{matsushita98}, and A 1060 \citep{tamura00}, and a mass
concentration within a radius 20--40 kpc has been recognized.

As for the metallicity distribution, significant radial gradients are
seen mainly in cD clusters, characterized by a metal concentration at
the center. It is discussed that the central excess in the metal
abundance is likely to be created by a bright central galaxy
\citep{degrandi04}. Since the metal accumulation, due to a type Ia
supernova and stellar mass loss, takes 2--10 Gyr, the central regions
need to be undisturbed for a long time \citep{bohringer04}. Using
recent high-resolution data, the abundance rise at the center has been
shown to often accompany a sharp drop right at the cluster center
within $r < 10$--$30$ kpc (Perseus: \cite{schmidt02,sanders04};
Centaurus: \cite{sanders02}; 0335+096: \cite{mazzotta03}; A 2052:
\cite{blanton03}; and A 2199: \cite{johnstone02}). A projection effect
of a multi-temperature gas is discussed as being a part of the cause
(\cite{molendi01}), while in some systems abundance depletion seems to
be a real feature.  A two-dimensional metallicity distribution was
first studied using the ASCA data for Centaurus cluster and
2A~0335+098, which indicated a patchy distribution of iron (e.g.\
\cite{tanaka03}). A Chandra observation of a cluster around 4C~+55.16
showed an unusual iron-rich plume \citep{iwasawa01}, and AWM~7
\citep{furusho03} revealed a blob-like metal concentration with a size
0.3--1 kpc near the center. Therefore, metal injection from the
central galaxy may take place in a highly inhomogeneous way.

To further understand the ICM properties and metal-injection process,
a detailed study of non-cD type clusters is useful. By comparing the
metallicity feature, we can separate the effect of cD galaxies and
estimate how central enrichment proceeded.  A nearby cluster of
galaxies, A 1060 ($z=0.0114$), is known to have no cool or sharp X-ray
peak at the center, and its proximity and brightness enable us to
evaluate the detailed temperature structure and iron distribution in
the central region. The cluster has 2 giant elliptical galaxies, whose
X-ray emissions were found to be very compact \citep{yamasaki02},
which helps us separate the ICM emission easily. Also, the temperature
in this cluster is very uniform with little sign of a cool component
\citep{furusho01}. This leaves little systematic ambiguity in
estimating the iron abundance simply from the equivalent width.

The Chandra results on the 2 giant elliptical galaxies (NGC~3311 and
NGC~3309) near the center of A 1060 were reported in
\citet{yamasaki02}. The present paper deals with the distributions of
the temperature and metallicity in the central ICM in some detail.
Preliminary results were given in \citet{furusho02}.  We assume $H_0 =
75$ km s$^{-1}$ Mpc$^{-1}$ with $q_0 = 0.5$, and an angular size of
$1'$ corresponding to 13 kpc. The solar number abundance of Fe relative
to H is taken as $4.68 \times 10^{-5}$ \citep{anders89}. We employ
Galactic absorption as $N_{\rm H} = 6\times 10^{20}$ cm$^{-2}$
throughout the paper. 

\section{Observation and Analysis} 

The Chandra observation of A 1060 was carried out on 2001 June 4, with
ACIS-I, for a total exposure time of 32 ks in the Very Faint Mode. The
CIAO ver 3.0 package was used in the data reduction and a calculation
of the energy response.  We generated background images and spectra
based on blank-sky data and software prepared by Maxim
Markevitch.\footnote{
http://hea-www.harvard.edu/\~{}maxim/axaf/acisbg/.} For a spectral
fitting, XSPEC ver 11.2.0 was used, and because of a known problem of
the ACIS response below 1 keV, we added the ACISABS absorption model
available in XSPEC\@.

\begin{figure}[!hb]
\centerline{\FigureFile(8cm,5cm){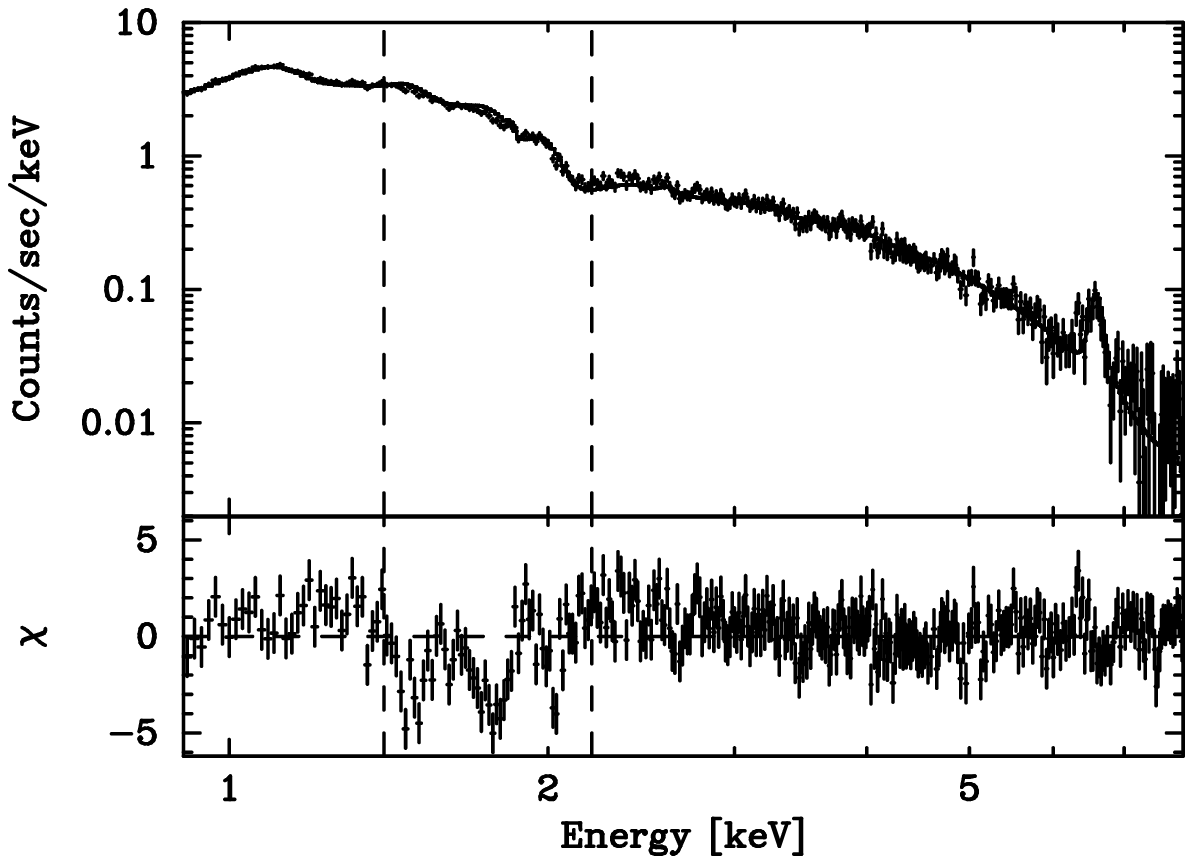}}
\caption{X-ray Spectrum of A 1060 for the sum of the 4 ACIS-I chips.
The data were fitted with a MEKAL thermal model with interstellar and
ACISABS absorptions. The bottom panel shows the residuals of the
fit. The model fit has a problem in the energy range 1.4-2.2 keV,
which was excluded in the spectral analysis. }
\label{acis_spec}
\end{figure}

Before proceeding to spatially resolved spectroscopy (subsection 4.3),
we fit the overall cluster spectrum to look at the consistency with
the previous results.  Figure \ref{acis_spec} shows the spectrum
between 0.9 and 8.0 keV from the entire ACIS-I chips fitted by a
single-temperature MEKAL model, including Galactic $N_{\rm H}$ and
ACISABS absorption.  A large deviation from the model, which might be
due to a calibration uncertainty of the Ir edge of the X-ray mirror,
is seen in the energy range 1.4--2.2 keV as being bounded by dashed
lines.  By masking this region, the fit was improved from
$\chi^{2}/\nu = 732/361$ to $\chi^{2}/\nu = 385/305$.  Therefore, we
do not include this region in a later spectral analysis. The
obtained spectral parameters for the temperature and metal abundance are
$kT = 3.36^{+0.05}_{-0.06}$ keV and $Z = 0.33^{+0.03}_{-0.03}$ solar,
respectively.  They are in reasonable agreement with the values from
EXOSAT ($ 3.3^{+0.4}_{-0.3}$ keV and $< 0.46$ solar by \cite{edge91})
and ASCA ($ 3.24\pm 0.06$ keV and $0.30\pm 0.02$ solar by
\cite{fukazawa98}).

\section{X-Ray Image}

\begin{figure}[!tb] 
\FigureFile(8cm,0cm){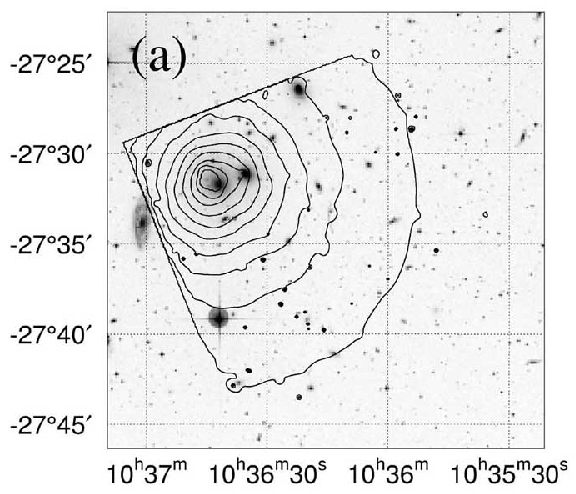}
\FigureFile(8cm,0cm){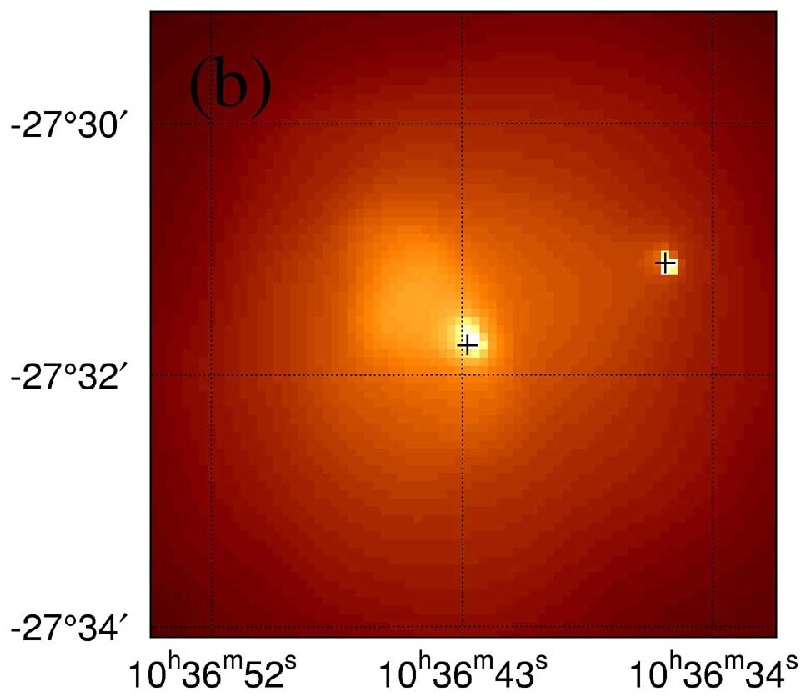}
\caption{ (a) Optical image of A 1060 taken from the Digitized Sky
Survey. The overlaid contours show a smoothed X-ray image.  (b)
Expanded X-ray image of the $5' \times 5'$ region centered on
A 1060. The crosses show the positions of NGC~3311 (center) and
NGC~3309 (north-west by $2'$), respectively.  }
\label{Xopt_image}
\end{figure}

The X-ray intensity contours obtained with ACIS-I are overlaid on the
optical image in figure \ref{Xopt_image}a. The data are for the energy
range 0.9--8.0 keV, and have been corrected for vignetting and
adaptively smoothed by a Gaussian function with a $1\sigma$ width of
$2''$.  Except for the 2 peaks corresponding to the bright central
galaxies, NGC 3311 and NGC 3309 as shown in \citet{yamasaki02}, the
ICM in A 1060 is very smooth and devoid of any prominent spatial
structures. This is in a marked contrast to the X-ray filaments and
cavities that are clearly recognized in the Chandra images of cD-type
clusters.

As can be seen in the expanded X-ray image in figure \ref{Xopt_image}b, 
there is a small extended emission apparently trailing in the
northeast direction from NGC 3311 with an angular scale $\lesssim 1'$.
We examined whether this excess feature might correspond to the
potential minimum of the cluster. The center of the cluster-wide X-ray
emission can be estimated from ROSAT data, which covers a diameter of
$\sim 2^\circ$. A two-dimensional fit for the ROSAT data, excluding the
central two galaxies, indicate that the center of the ICM emission is
$28''$ southwest of NGC 3311, opposite to the direction of the excess
emission. This suggests that the excess X-ray emission could be
related to the elliptical galaxy NGC 3311. However, since many
clusters show their X-ray peaks offset from the isophotal centroid,
there remains a possibility that the feature in the northwest of
NGC~3311 corresponds to the minimum of the gravitational potential.

Figure \ref{Xopt_image}b also shows a weak source at $30''$ south of
NGC 3311, which is inside the isophotal radius ($95''$) of NGC 3311\@.
Although the nearest known source is a globular cluster SGM95 33,
it is $2''$ offset from the X-ray position, and is unlikely to be
the counterpart.

\section{Radial Profiles}

\subsection{Temperature and Metal Abundance}

\begin{figure}[!tb]
\centerline{\FigureFile(8cm,5cm){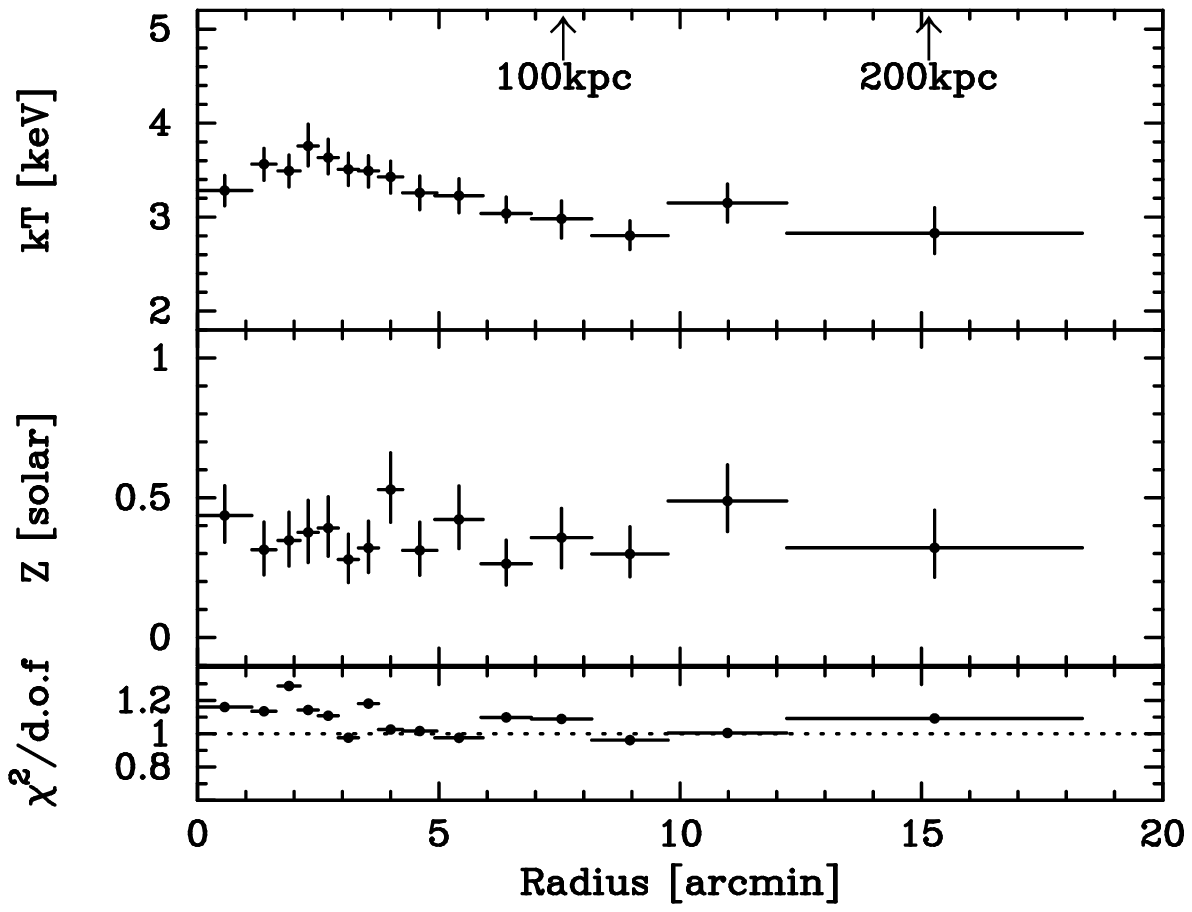}}
\caption{Radial distribution of the temperature (top), metal
abundance (middle), and the $\chi^2$ value of the spectral fit (bottom),
based on a spectral fit for concentric annular regions. The errors are
single-parameter 90\% limits.}
\label{ring_fit}
\end{figure}

The radial distributions of the ICM temperature and metallicity were
examined. The two central galaxies (for NGC~3311, $r\sim 20''$, for
NGC~3311, $r\sim 15''$) and other point sources (twice as large as the
detected radius based on CIAO wavelet and cell detection tools) were
excluded.  The annular spectral data for radii $r > 20''$ were
accumulated in concentric annular regions by keeping the number of
photons greater than 10000 in each annulus.  Figure \ref{ring_fit}
shows the radial distributions of the temperature, metal abundance,
and the $\chi^2$ value for the spectral fit by absorbed MEKAL models.
The bottom panel indicates that a single-temperature thermal model
gives a reasonably good fit with reduced $\chi^2$ values less than 1.4
in all of the regions.

As can be seen in figure \ref{ring_fit}, the temperature shows a rise
up to about 3.7 keV at $r = 2'$--$3'$ from the center. This feature is
significant, and a hypothesis of a constant temperature over the whole
radius is rejected with more than the 90\% confidence.  The highest
temperature ($\sim 3.7$ keV) exceeds the level in the outer region by
about 20\%.  In the X-ray image, however, there is no clear structure
or edge-like feature at this radius.  We also note that the central
temperature of about 3.4 keV is much higher than those seen in
so-called cooling flow clusters, which generally show a temperature
drop down to $1/2$--$1/3$ of the outer level. Therefore, gas cooling
is certainly not a major process occurring at the center of
A 1060\@. In the outward direction, the temperature gradually declines
from $r = 2'$ to $8'$ by more than 0.5 keV in a monotonic way. Since
the ICM temperature in the outer region ($r\sim 30'$) measured with
ASCA is about 3.0 keV \citep{furusho01}, the temperature gradient is
confined only in the central region ($r<8'$ or 100 kpc).

The metal abundance in figure \ref{ring_fit} is consistent with the
overall value of 0.33 solar in all regions.  The innermost region
($r<1'$) suggests a somewhat high value of $0.44^{+0.11}_{-0.10}$
solar, but no systematic rise toward the center, as seen in other
cD-type clusters, is recognized. The constant metallicity and the
observed value are consistent with the previous ASCA result
\citep{furusho01,tamura00}.

\subsection{Density Profile}

\begin{figure}[!tb]
\centerline{\FigureFile(8cm,5cm){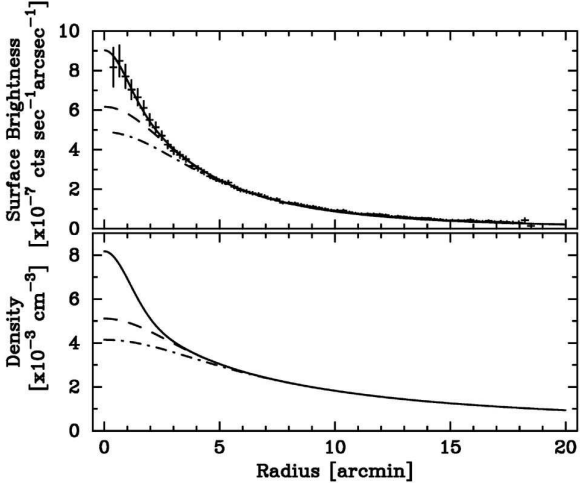}}
\caption{Radial distribution of the surface brightness (top) and
density (bottom). The data points are indicated by crosses, in which
point sources and the central two galaxies are subtracted. The model
approaches the observed data by iteratively enhancing the central
density. The 3 lines correspond to the initial ($\beta$ model,
dash-doted line), the 2nd (dashed line), and the 7th (solid line)
iterations, respectively. }
\label{iteration}
\end{figure}

We plot the surface brightness profile in the energy range 0.9--8.0
keV in figure \ref{iteration}.  The blank-sky data were subtracted in
this plot.  The data were fitted with a single $\beta$ model, but the
fit is not acceptable because of a discrepancy in the central
region. The data seem to require an additional narrow
component. Although a double $\beta$ model is often used to fit the
data, the superposition of 2 separate emission components is
physically difficult to interpret. We explored a solution of a
3-dimensional density structure with a single emission component based
on the assumption that the temperature and metallicity are constant with
radius, and also that the cluster is spherically symmetric, which seem to be
a reasonable approximation of the observed features.

We first excluded the central region, and only fitted data in the
outer region ($r = 4'$--$18'$), which gave an acceptable fit (the
bottom dash-dot curve in figure \ref{iteration}). The fitted
parameters are $r_c = 5.5'^{+1.4'}_{-1.3'}$ and $\beta =
0.56^{+0.07}_{-0.05}$, respectively. We then tried to modify the
radial density profile within $r \le 4'$ from the $\beta$ model, and
found an acceptable one that gives a surface brightness profile
consistent with the observed data. To find the best estimate of the
density profile, we performed an iteration in the following way.

We took the ratio between the surface brightness ($\Sigma$) data and
the single $\beta$ model, which is about 1.8 at the center, and drops
to unity at $r = 4'$. Since the emissivity scales as the square of the
density, we took the square root of the brightness ratio, i.e.\ $\eta
= \sqrt{\Sigma_{\rm data}/\Sigma_{\beta}}$, which varies as a function
of the radius, and used it as a correction factor to the $\beta$-model
density. The $\beta$-model density profile was multiplied by $\eta$,
and the predicted brightness profile was compared with the data.  In
this process, the ratio as a function of the radius was approximated
by the sum of a Gaussian function and a constant, which was found to
give a good fit to the observed profile.

The brightness and density profiles at this stage are shown by the 2nd
curves from the bottom (dashed) in the top and bottom panels of figure
\ref{iteration}.  The resultant brightness profile became steeper than
the pure $\beta$ case, but not as sharp as the data, because our
correction factor is ``diluted'' by the emission in the cluster
outskirts superposed in the line of sight. This corrected $\beta$
model was replaced by the previous $\beta$ model, and we took the
ratio of the surface brightness again and calculated further
correction factors in the same manner for each radius. This iteration
process was ended when the brightness profile gave an acceptable
$\chi^2$ value to fit the data.  This took 7 iteration steps. T
resultant profile is shown in the top curve (solid) of figure
\ref{iteration}. The 3-dimensional density profile now shows a fairly
sharp peak in the center, with a density about twice as much as that
of the $\beta$ model (top solid curve in the bottom panel of figure
\ref{iteration}). We use this density profile in a later estimation of
the physical parameters in the core region.  The central electron
density estimated here is
\[
 n_0 = 8.2^{+1.8}_{-1.0} \times 10^{-3}\ {\rm cm}^{-3},
\]
which is still lower than the level in other cD-type clusters.

\section{Spatial Distribution}
\subsection{Temperature Distribution}

ASCA observations have shown that the ICM of A 1060 was very isothermal
\citep{tamura00,furusho01}. However, the angular resolution of ASCA
did not allow us to look into the temperature structure on spatial
scales less than a few arcmin.  Here, we examine the temperature
structure within $r<20'$ from the center based on the Chandra data.
The hardness ratio (HR) was calculated for two energy bands, 2--6 keV
and 0.9--2 keV, after a point-source elimination and background
subtraction. Regions containing NGC~3311 and NGC~3309 were masked out
with a diameter of $1'$. We assumed a single-temperature MEKAL model
with a metal abundance of 0.3 solar absorbed with Galactic $N_{\rm H}$
and ACISABS\@.

\begin{figure}[!tb]
\centerline{\FigureFile(8cm,5cm){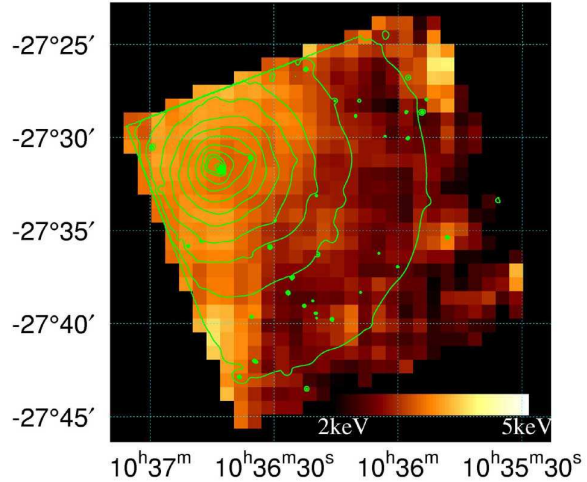}}
\caption{Temperature map obtained from the hardness ratios. The pixel
size is $44''\times 44''$ (10 kpc across). The color coding was
performed for a temperature range $kT=2$--$5$ keV\@. The X-ray
intensity contours are the same as in figure \ref{Xopt_image}a.}
\label{kt_map_hr}
\end{figure}

The absence of an intense soft component and the constancy of the
metal abundance in A 1060 assures us that the HR values are reliable
measures of the temperature.  The effect of a metallicity variation on
the derived HR values is fairly small.  A change of the metal
abundance from 0.3 to 0.5 solar in the model results in a temperature
shift of 0.06 keV (less than the statistical error of about 0.1 keV)
from the nominal level of 3 keV\@.

Figure \ref{kt_map_hr} shows a color-coded temperature map obtained
from the HR distribution.  The statistical error in the temperature is
typically 0.3~keV ($1\sigma$) at $r\sim5'$.  The pixel size is $44''
\times 44''$ (10 kpc across), and we took a running average for the HR
over $4 \times 4$ pixels. The map confirms the general trend that the
temperature in the outer region is cooler than the cluster
center. Also, the region around $r \sim 3'$ appears to be hotter than
at the exact center, which is consistent with the radial temperature
profile in figure \ref{ring_fit}. We also note that the warm ring at
$r \sim 3'$ shows no clear azimuthal structure, and that the variation
of the temperature looks fairly gradual in all regions.

\begin{figure}[!tb] 
\FigureFile(8cm,5cm){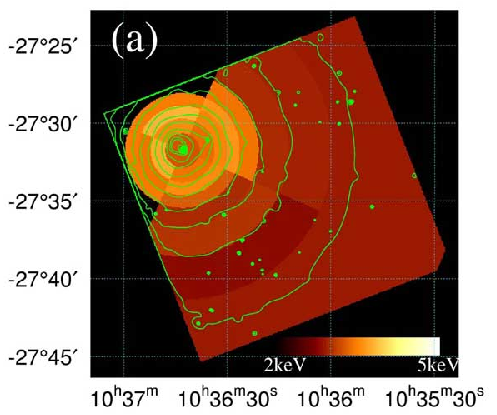}
\FigureFile(8cm,5cm){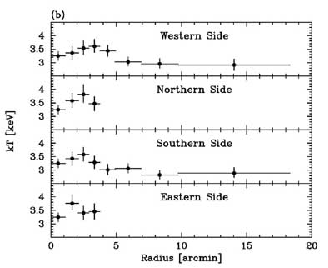}
\caption{(a) Temperature map obtained from spectral fits.  (b) Radial
temperature profiles for 4 directions, based on the results shown in
(a)}
\label{kt_map_fit}
 \end{figure}

To confirm the temperature distribution indicated by the hardness
ratio analysis, we performed spectral fits by accumulating data in
larger regions. For 8 radially divided annuli, we further limit their
opening angles to $90^\circ$. The spectral data were accumulated,
background-subtracted, and fitted with a single-temperature MEKAL
model. The absorption was fixed to the Galactic value, and the metal
abundance fixed to 0.3 solar.  We thus obtained the temperature map
shown in figure \ref{kt_map_fit}a. The similarity with figure
\ref{kt_map_hr} is quite evident, and the result confirms the
relatively small variation of temperature in this cluster. The
temperature in the outermost ($r> 10'$) large region is 2.9 keV, which
is significantly cooler than the level in the central region.  The
temperature values are also plotted in figure \ref{kt_map_fit}b as a
function of the radius for 4 different azimuthal directions. Here, we
confirm that the temperature shows a broad peak at $r \sim 3'$ in all
directions. There is no significant difference in the 4 profiles.

\subsection{Metal Distribution}

The isothermal ICM of A 1060 helps us look into the metallicity (iron
in particular) distribution in a straightforward way. Since the
equivalent width of the Fe-K line changes by about 10\% for a
temperature change of 1 keV at $kT \sim 3$ keV, the constant
temperature ($\Delta T < 0.3$ keV) suppresses the systematic ambiguity
in determining the metallicity. To estimate the Fe abundance
distribution, we took the HR between energy bands 6--7 keV and 2.2--6
keV\@. The hard band contains the Fe-K emission line at 6.7 keV
(redshifted to 6.62 keV), and the fraction of the line photon is about
10\% for a 3 keV plasma containing 0.3 solar abundance of iron. The
soft band is devoid of any prominent line, such as those of Si, and
the intensity is a good measure of the continuum level. In fact,
changing the metal abundance from 0.3 to 0.5 solar at $kT = 3$ keV
pushes up the 2--6 keV flux by only $\sim 6$\%. These features
indicate that the HR for these 2 energy bands is a good measure of the
iron abundance.

\begin{figure}[!tb] 
\FigureFile(8cm,5cm){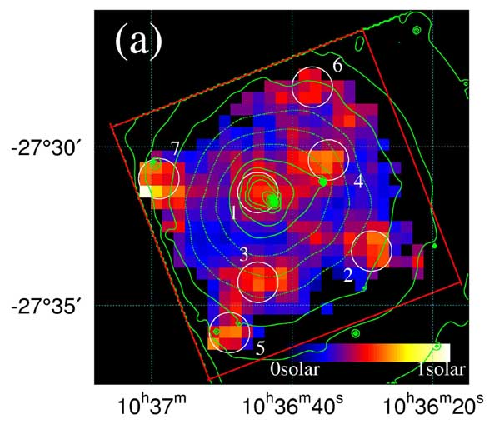}
\FigureFile(8cm,5cm){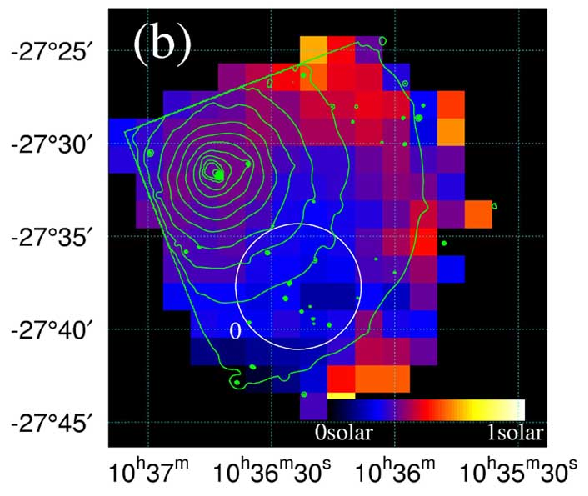}
\caption{(a) Distribution of the iron-band hardness ratios, smoothed
in $22'' \times 22''$ pixels (5 kpc across). Results for only ACIS-I3
chip (indicated with red square) are shown. Several high-abundance
regions are encircled and numbered. The contours indicate the X-ray
intensity.  (b) Same distribution as in (a), but the data were
smoothed in $88'' \times 88''$ pixels (20 kpc across) to show a larger
scale distribution. The extended low-abundance region is numbered as 0
and encircled.}
\label{ab_map_hr}
\end{figure}

The ratios were calculated for the background-subtracted data. The
spatial distribution of the HRs are shown in figure \ref{ab_map_hr}a as
a color-coded map, in which the data are accumulated in $22''\times
22''$ pixels. In this plot, the HR values were converted to iron
abundance assuming a constant temperature of $kT = 3.0$ keV\@.
The statistical error on the abundance is typically $\pm 0.1$ solar
everywhere.
We notice that there are
some regions with high metallicity in this hardness map. These regions
are numbered and marked with circles. In the HR values, their
statistical significances are between $1.7\sigma$ and $2.5\sigma$. Since
the number of pixels in $r<5'$ is about 500, it is unlikely that these
deviations are due to statistical fluctuation.

To raise the statistical quality of the HRs in the outer regions, the
data were further binned in $88'' \times 88''$ pixels, and the
corresponding temperatures are plotted in figure
\ref{ab_map_hr}b. There are some high abundance regions seen in the
northwest edge. The statistical significance is $\sim 2.5\sigma$.
Apart from these high-abundance patches near the field edge, the
values seem to be fairly constant in a wide region with $r = 3'$--$10'$.

\begin{figure}[!tb]
\FigureFile(8cm,5cm){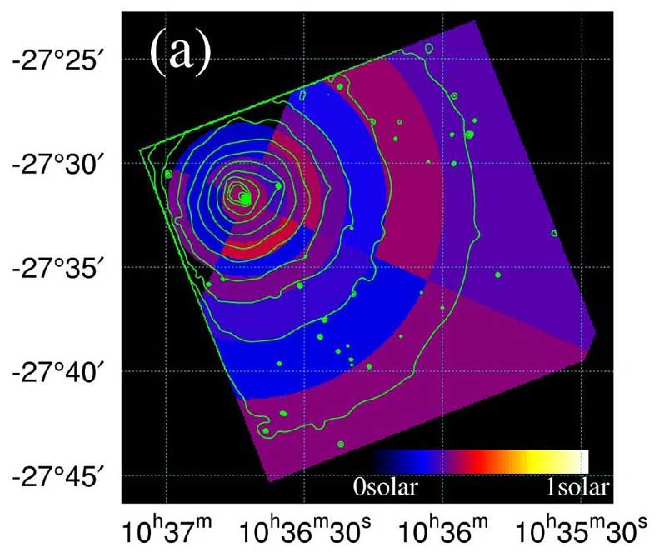}
\FigureFile(8cm,5cm){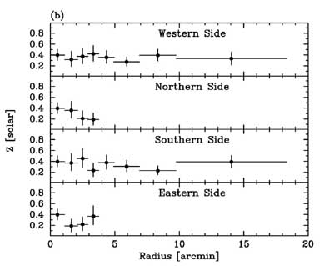}
\caption{(a) Spatial distribution of metal abundance based on spectral
fits.  (b) Radial abundance profiles for 4 directions, obtained from
the spectral fits as  shown in panel a.}
\label{ab_map_fit}
\end{figure}

Again, we carried out a spectral analysis for the segmented regions,
and the results on the metal abundance are shown in figure
\ref{ab_map_fit}a. Even though most of the notable structures are
smoothed out, the results are generally consistent with the HR results
shown in figures \ref{ab_map_hr}a and b. We also note that there is no
systematic drop or rise as a function of the radius in the abundance
feature. The same results are plotted in figure \ref{ab_map_fit}b
as a function of the radius.

\section{High Metallicity Regions}

\begin{figure}[!tb] 
\centerline{\FigureFile(8cm,5cm){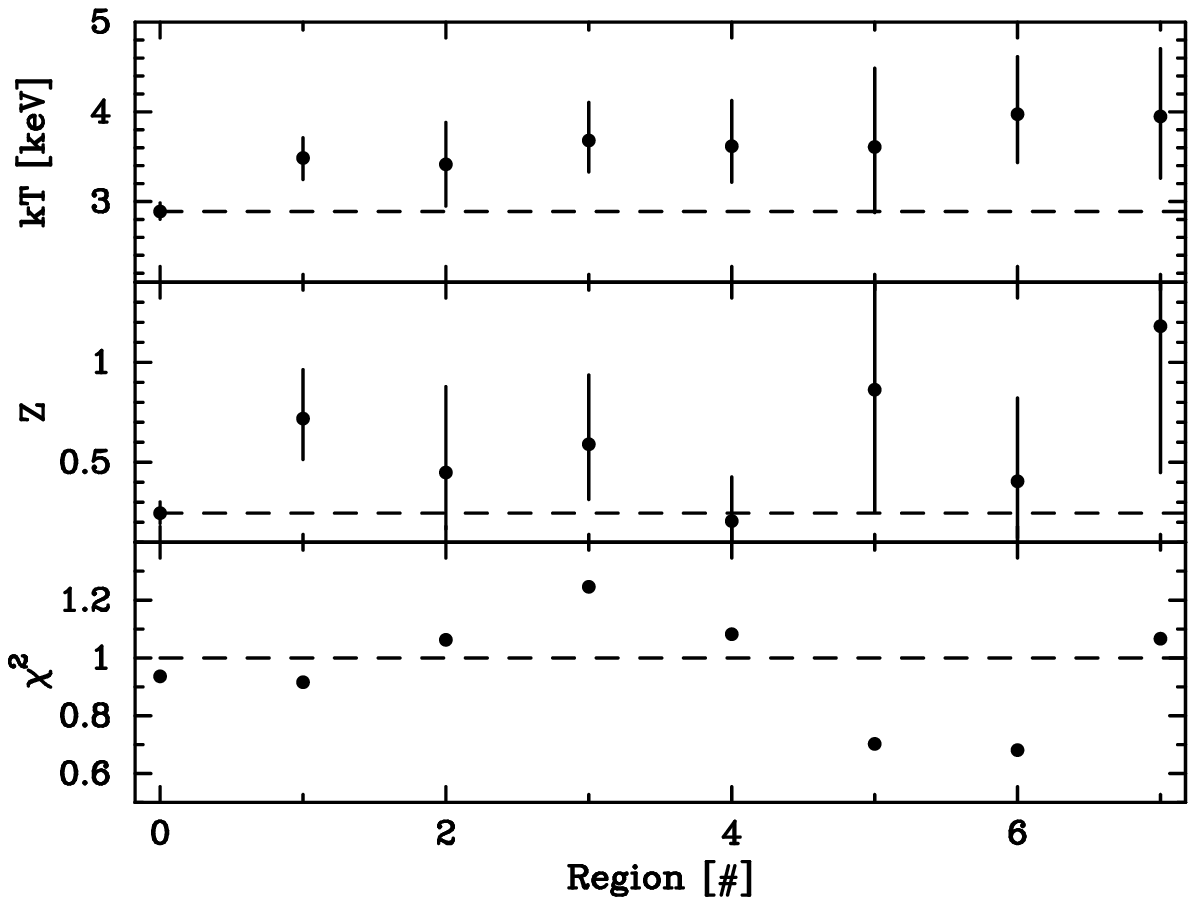}}
\caption{Temperature and metal abundance obtained from spectral
fits for the regions marked in figures \ref{ab_map_hr}a and b. The
errors are single-parameter 90\% limits.}
\label{highZfits}
\end{figure}

As shown in the previous section, there are several high-metallicity
regions near the center of the A 1060 cluster.  Here, we evaluate their
properties based on the spectral fit.  The pulse-height data were
accumulated separately for the 8 regions indicated in figures
\ref{ab_map_hr}a and b, and spectral fits were performed. Table
\ref{tab:first} summarizes the fit, and the results are plotted in
figure \ref{highZfits}.  Data \#0 is the extended low-abundance
region, and \#1--7 are the high-abundance ones. The errors are
single-parameter 90\% limits ($1.65 \sigma$). Compared with the \#0
level, most of the regions indicate deviations by less than $2\sigma$,
and the significances are rather low. Only region \#1 shows a $3.4
\sigma$ deviation, which strongly suggests that Fe is concentrated in
this small region with a diameter less than 10 kpc. We refer to this
region as ``metal-rich blob'' hereafter. We also note that the
temperature of regions \#1 is not significantly different from that of
the surrounding level.

\begin{table}[!ht] 
  \caption{Results of spectral fits for selected
 regions in figure \ref{ab_map_hr}.}\label{tab:first}
  \begin{center} \scriptsize
    \begin{tabular}{ccccc}  \hline \hline
      Region & $kT$    & $Z$  & $F_{\rm X}$ & $L_{\rm X}$ \\ 
        (\#)  & (keV) & (solar)  & (erg cm$^{-2}$s$^{-1}$) &
     (erg s$^{-1}$) \\ \hline
      0    & $2.95^{+0.12}_{-0.09}$ & $0.26^{+0.05}_{-0.06}$ &
     $6.90^{+0.07}_{-0.11} \times 10^{-12}$ & $1.73^{+0.02}_{-0.03} \times 10^{42}$  \\ 
      1    & $3.55^{+0.25}_{-0.25}$ & $0.74^{+0.29}_{-0.23}$ &
     $1.10^{+0.05}_{-0.07} \times 10^{-12}$ & $2.75^{+0.01}_{-0.02} \times 10^{41}$ \\ 
      2    & $3.51^{+0.57}_{-0.48}$ & $0.44^{+0.53}_{-0.31}$ &
     $4.46^{+0.29}_{-0.49} \times 10^{-13}$ & $1.12^{+0.07}_{-0.12} \times 10^{41}$  \\ 
      3    & $3.85^{+0.48}_{-0.37}$ & $0.77^{+0.50}_{-0.36}$ &
     $6.40^{+0.34}_{-0.72} \times 10^{-13}$ & $1.60^{+0.09}_{-0.18} \times 10^{41}$ \\ 
      4    & $3.70^{+0.57}_{-0.41}$ & $0.29^{+0.31}_{-0.21}$ &
     $7.07^{+0.37}_{-0.53} \times 10^{-13}$ & $1.77^{+0.09}_{-0.13} \times 10^{41}$  \\ 
      5    & $3.20^{+1.84}_{-0.77}$ & $0.44^{+0.32}_{-0.42}$ &
     $1.60^{+0.25}_{-0.81} \times 10^{-13}$ & $4.00^{+0.63}_{-2.03} \times 10^{40}$ \\ 
      6    & $4.00^{+0.65}_{-0.71}$ & $0.38^{+0.48}_{-0.31}$ &
     $2.58^{+0.28}_{-0.49} \times 10^{-13}$ & $6.46^{+0.70}_{-1.23} \times 10^{40}$  \\ 
      7    & $3.66^{+1.06}_{-0.72}$ & $0.65^{+1.68}_{-0.52}$ &
     $4.46^{+0.23}_{-0.46} \times 10^{-13}$ & $1.12^{+0.06}_{-0.12} \times 10^{41}$ \\  \hline 
    \end{tabular}
  \end{center}
\end{table}

\begin{figure}[!tb]
\centerline{\FigureFile(8cm,5cm){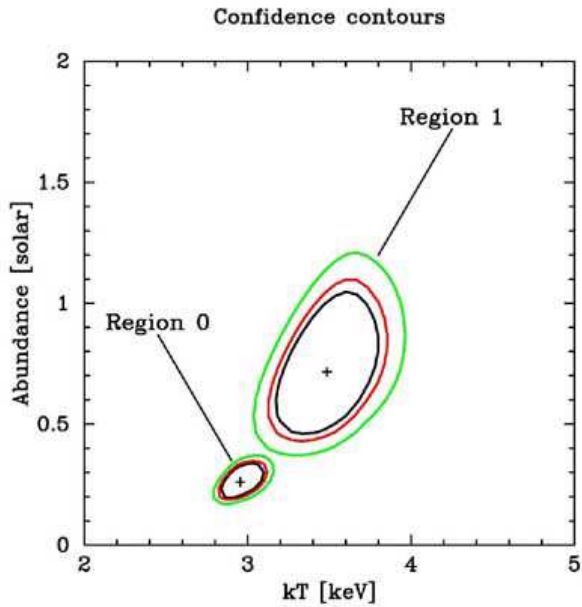}}
\caption{Confidence contours for the spectral parameters for the
pulse-height spectra of regions \#0 and \#1. The confidence levels are
90\% (black), 95\% (red), and 99\% (green) for 2 interesting
parameters. The two spectra have different abundance at more than the
99\% confidence level.}
\label{contour}
\end{figure}

To make a close spectral comparison between regions \#0 and \#1, we
show confidence contours of the fit in figure \ref{contour}.  The
metal abundance in \#1 is 0.74 solar, which is about 3 times higher
than the \#0 level of 0.26 solar. The \#1 temperature is higher than
\#0 by 0.5 keV, but about the same as in the surrounding regions.  The
2-dimensional confidence contours indicate that these 2 spectra have
a different abundance at the 99\% confidence level.

To compare with the iron abundance feature, we consider the optical
galaxy distribution, which is shown in figure \ref{Xopt_image}a. The
abundance enhancement around the north of NGC 3311 may be related to
this galaxy (or an internal structure of the galaxy), since the
isophotal radius of NGC 3311 ($r_{\rm e} = 95''$) is outside of the
high-abundance region. Another iron concentration is seen at $3'$ to
the southwest of the cluster center (RA $\simeq 10^{\rm h}36^{\rm
m}28^{\rm s}$, Dec $\simeq - 27^\circ 33'$). However, no galaxies are
seen in the optical image. These features indicate that the iron
enhancement does not generally occur at the positions of the present
bright galaxies. This suggests that the injection of iron into the ICM
may have occurred in the past when the galaxy positions and
brightnesses were much different from the present configuration.

\section{Discussion} 
The present analysis on the Chandra data of A 1060 confirmed the
absence of a central cool component with high spatial resolution. At
the same time, there are significant inhomogeneities in the spatial
distributions of the temperature and metallicity. We discuss their
implications in this section.

\subsection{Temperature Distribution}

First, let us briefly examine the temperature structure. The radial
temperature profile shows a peak at around $\sim35$~kpc ($\sim160''$)
from the center with a level 20\% higher than that of the outer
region, accompanied by a gradual decline toward the center. The
temperature maps and the profiles in 4 directions (figures
\ref{kt_map_hr} and \ref{kt_map_fit}) indicate that this
high-temperature region is encircling the central region ($r<5'$) with
no sharp difference in the azimuthal directions. Previous ASCA
observations did not have the necessary spatial resolution to resolve
these features \citep{tamura00, furusho01}.

Numerical simulations of cluster formation generally predict a fairly
steep temperature drop toward the outer regions \citep{frenk99}, with
a factor of 2 drop at a radius of $\sim 0.5\ r_{180}$. Here, $r_{180}$
is the radius within which the mean density is 180 times the critical
density, and the matter is approximately virialized within this
radius. For A 1060, $r_{180} \approx 1.5$ Mpc, which is 20 times larger
than the radius for the observed temperature drop. Therefore, the
``hot ring'' is likely to be caused by a process taking place only in
the central region. Since the ICM temperature is fairly constant at
3.0 keV for $r = 8'$--$30'$ as shown by the present data and the previous
ASCA results \citep{furusho01}, we should think that extra energy is
carried in within the central 50 kpc.

A 1060 has 2 bright central galaxies, NGC~3311 and NGC~3309, near the
center, separated by 22 kpc in projection. This suggests that A 1060
may have gone through a merger in the long past, with the above 2
galaxies located at the centers of each component. During the merger,
the central galaxies may have lost a large portion of the halo gas,
but the stellar component would have remained. This can explain the
observed very compact X-ray halos of these 2 galaxies
\citep{yamasaki02}. Since the 3-dimensional distance between the 2
galaxies should be larger than 22 kpc, the diameter of the ``hot
ring'' is comparable to the region covered by the motions of the
central galaxies. However, it is not obvious how extra energies were
created in this region. Merger should create a temperature structure,
such as that seen in the double-core cluster, A 2256. However, because
the ICM of A 1060 is much more isothermal, the temperature variation
directly caused by the merger is likely to have already been smoothed
out.

One possibility that we consider here is the energy input from bright
galaxies. Supernova and stellar winds in the galaxies can inject a
significant amount of heat into the surrounding ICM in the form of
galactic winds. We can perform a crude estimation by considering only
the supernova effects. Based on the supernova rate, $(0.16\pm
0.06)\times 10^{-12}$ yr$^{-1}L_{B\odot}^{-1}$, given by
\citet{cappellaro99}, the total supernova rate in the 2 elliptical
galaxies ($L_B \approx 1.9\times 10^{11} L_\odot$) is estimated to be
$\sim 0.03$ yr$^{-1}$.  Assuming that each supernovae gives an energy
of $10^{51}$ erg, the total energy input is $3\times 10^{58}$ erg in
$10^9$ yr, which is roughly 20\% of the thermal energy contained in a
volume with the radius of 25 kpc in the cluster center.  This
estimation suggests that, over a time scale of $10^9$ yr, energy input
from galaxies can make a significant contribution to the thermal
structure of the ICM\@.  We, however, note that because the central
galaxies contain cool gas compared with the ICM \citep{yamasaki02},
either a high galaxy motion or a wind-like mass ejection would be
necessary to heat the surrounding ICM efficiently.

The slight drop of temperature at the center may be explained by
cooling.  The radiative cooling time at the center of A 1060 is
estimated to be $6.6\times 10^{9}$ yr, which is comparable to the
Hubble time ($1.3\times 10^{10}$ yr), and much longer than the heat
conduction time shown below.  Therefore, radiative cooling at the
cluster center is not a dominant effect.  The heat-conduction time
scale can be calculated from the central density and core radius given
by the $\beta$-model fit.
The conduction time at the cluster center is estimated to be
$1.4\times 10^9$ yr, which is ten times shorter than the Hubble time.
The observed isothermality, better than most of other clusters, in the
central region of A 1060 suggest that the heat conduction is working at
the theoretical level, and that the suppressing effect, such as due to
a magnetic field, is small.

\subsection{Mass Distribution}

\begin{figure}[!tb]
\FigureFile(8cm,5cm){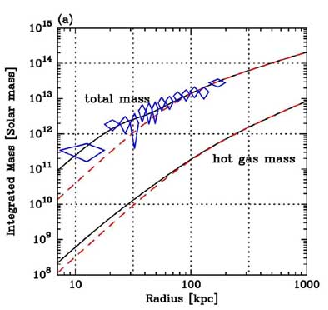}
\FigureFile(8cm,5cm){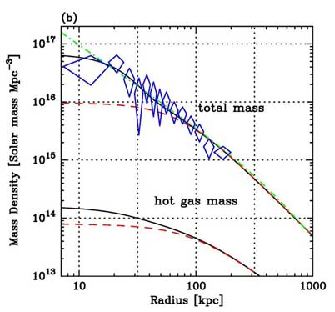}
\caption{(a) Gravitational mass profile estimated from the simple
$\beta$-model (upper dashed line) and the density profile shown in
figure \ref{iteration} (upper solid line). The lower lines represent
the ICM mass based on the $\beta$-model and the density profile in
figure \ref{iteration}, respectively. The diamonds show the results
when the temperature variation shown in figure \ref{ring_fit} is
considered.  (b) Same as (a), but for the differential mass density
profiles.  The dash-dotted line shows a NFW model with a central
density slope of 1.5.}
\label{grav_mass}

\end{figure}

Based on an approximation of constant temperature and metallicity as
well as spherical symmetry, we have derived the density distribution
in a model-independent way. Compared with the simple $\beta$ model
that can fit the data at $r>4'$, the central gas density is about
twice higher, as shown in figure \ref{iteration}. If the ICM is in
hydrostatic equilibrium, the observed high density of the gas simply
indicates a concentration of the gravitational mass. Assuming an
isothermality for simplicity, we have derived a gravitational mass
profile, which is shown along with the gas mass profile in figure
\ref{grav_mass}a. The corresponding mass density profiles are also
shown in figure \ref{grav_mass}b.  In these figures, the mass profiles
for a simple $\beta$ model are plotted for a comparison. We notice an
excess in the central region within $r = 50$ kpc over the simple
$\beta$ profile, and the gravitational mass is about 5 times higher in
the central 10 kpc.  Since this feature may have resulted due to our
neglect of the temperature variation shown in figure \ref{ring_fit},
we also plot a mass curve including the temperature variation.  The
results are shown by diamonds in figures \ref{grav_mass}a and b.

\citet{tamura00} showed, based on the analyses of the ASCA and ROSAT
data of A 1060, that the mass profile is well described by a NFW
profile \citep{navarro97} with a central slope $\sim 1.5$. Their
analyses, however, did not exclude the compact halo of NGC~3311, which
may have acted in favor of the cusp-like central structure.

The present result confirms that there is a significant concentration
of dark matter in the central 50 kpc region, and that baryons (gas and
galaxies) do not simply follow this feature. In fact, the radial
profile of the baryon fraction shows a marked drop at the
center. \citet{ota02} reported, based on the analysis of 79 cluster
data obtained by ROSAT, a double-peak distribution for the core radius
with peaks at 60 kpc and 220 kpc. Therefore, these observed results
jointly suggest that dark matter may preferentially be accumulated
within a radius of 50--60 kpc.

\subsection{Metal Distribution}

The Chandra data confirmed that the radial distribution of the metal
abundance is fairly uniform with an average value of $\sim0.3$ solar,
which agrees with the previous results \citep{tamura96}.  Recent
Chandra and XMM-Newton observations have revealed that in several clusters
of galaxies the radial metal abundance profile shows a peak offset
from the center.  \citet{furusho03} found that, in the poor cluster
AWM~7, this feature is mainly caused by two high-metallicity blobs
symmetrically placed around the center. The radial metallicity
distribution in A 1060 seems to be more uniform than that of the above
clusters. \citet{morris03} discussed that some abundance profile may
be related to the evolution and metallicity of the central cool
component. The flat radial distribution of metallicity in A 1060
suggests that this cluster is still in an early phase of evolution.

The present Chandra observation showed that the iron distribution in
the central region of A 1060 is not completely uniform. Several local
concentrations with roughly arcminute scales ($1'$ or $13$ kpc) are
recognized, with no spatial correspondence with the bright
galaxies. We here examine the origin of iron in the ICM and the
history of metal injection.

Let us first estimate the mass of iron in the high-metallicity
region. We consider only the region at the northeast of NGC 3311,
since it is the brightest and the most significant among the
high-metallicity regions. Because the measured abundance is
$0.74^{+0.29}_{-0.23}$ solar for the projected data, the true
abundance of the high-metallicity region must be significantly
higher. For this estimation, we need to subtract the foreground and
background components of the ICM in the line of sight.  We employ our
density profile shown in figure \ref{iteration}, which gives the
gas density at 10~kpc from the cluster center to be $7.6 \times 10^{-3}$
cm$^{-3}$.

Since the metal-rich region is more luminous than the ambient region
at the same distance from the cluster center, it must have a higher
density.  In the later estimation of the physical parameters, we
simply approximate that the region has a spherical blob shape with a
radius of 8.7 kpc ($40''$) filled with a constant gas density. The
measured value of the excess luminosity, $L_{\rm X}=
2.75^{+0.01}_{-0.02} \times 10^{41}$~erg s$^{-1}$, gives the gas
density to be about $1.1 \times 10^{-2}$ cm$^{-3}$ and the blob mass
about $7\times 10^8 M_\odot$. We note that the derived blob mass is
larger than that of the compact X-ray halo of NGC 3311, which is
estimated to be $5.6\times 10^7 M_\odot$ \citep{yamasaki02}. The
spectral fit indicates that the metal abundance of the blob region is
$\sim 0.7$ solar.  Again, this is a ``diluted'' value, since the data
contain all of the emission in the line of sight.  We corrected for
the projected effect, assuming the same density distribution. The
metal abundance of the blob is found to be 1.5 solar, which is 4--5
times higher than the average cluster value.

Based on these results, the total mass of Fe ($M_{\rm Fe}$) contained
in the blob is estimated to be $3.0 \times 10^6 M_\odot$. The question
is whether this much Fe can be supplied by a single galaxy,
NGC~3311. The stellar mass of NGC~3311 is estimated to be $4.8 \times
10^{11} M_\odot$, assuming $M/L_B = 8.5(M_\odot/L_\odot)$ and $L_B =
11.15$ \citep{faber89}. The X-ray halo of NGC~3311 gives a negligible
contribution. If the average Fe abundance in the stellar mass is 1
solar, we obtain $M_{\rm Fe} \sim 1.3\times 10^9 M_\odot$ as the total
stellar Fe content. The above-estimated $M_{\rm Fe}$ in the blob is
only 0.2\% of the stellar Fe content, and can be supplied by a single
galaxy. The metal-rich blob can possibly be a stripped halo of
NGC~3311, as far as the gas mass and Fe content are concerned.

To further examine the consistency of this picture, we estimate the
ram pressure stripping time for a galaxy moving in the ICM based on
the following formula \citep{sarazin88}:
\[
t \sim 3.0\times 10^7 \left( \frac{\rho_{\rm ISM}}{\rho_{\rm
ICM}}\right)^{1/2} \left( \frac{v}{10^3 {\rm km\
s^{-1}}}\right)^{-1}\left(\frac{R}{10\ {\rm kpc}}\right) \ {\rm yr}
\]
Taking $\rho_{\rm ISM}/\rho_{\rm ICM} \approx 10$, $v \approx 500$ km
s$^{-1}$, and $R \approx 2$ kpc \citep{yamasaki02}, the stripping time
was estimated to be about $1.0\times 10^7$ yr. The galaxy can travel
about 6~kpc within the stripping time, which is about the same distance
from NGC~3311 to the blob center.  Based on these supporting features, we
consider that the metal-rich blob is likely to be the gas stripped off
from NGC~3311.

Except for these blob-like metal concentrations, the general cluster
space seems to show a fairly uniform metal abundance. Therefore, we
need to understand how metals are eventually mixed into the general
ICM space. As pointed out by several authors
\citep{metzler94,ezawa97}, the diffusion of Fe in the ICM is very
slow, and mixing due to galaxy motions or to subcluster mergers would
be necessary. We hope that future observational and theoretical
studies will give us a comprehensive view on the metal enrichment
process.  An XMM-Newton observation of A 1060 will soon reveal the
metal distribution and temperature structure in more detail with its
larger collecting area.

\section{Conclusion}

A Chandra observation of the A 1060 cluster confirmed that the
temperature and metal abundance distributions are fairly uniform from
the outer region ($r \sim 230$ kpc) to the central 5 kpc region.
Regarding the temperature, the radial profile shows a broad peak at
around 35~kpc from the center at a level 20\% higher than that of the
outer region.  This may reflect the condition of the hydrostatic
equilibrium as numerical simulations suggest. These results suggest
that this cluster is relaxed, and has a relatively slow growth speed
of its evolution. Assuming a constant temperature and metal abundance,
we derived a 3-dimensional density profile by iteration on the initial
$\beta$ model. The gas distribution implies a concentration of
gravitational matter within a radius of 50 kpc. For metal abundance,
we detected several metal-rich regions around the central region.  In
particular, the central blob adjacent to the galaxy NGC 3311 exhibits
a high abundance (1.5~solar) and gas density ($1.1 \times
10^{-2}$~cm$^{-3}$). We interpret this to be caused by gas-stripping
from NGC~3311.

\bigskip T.F. acknowledges support from the Japan Society for the
Promotion of Science. This work was partly supported by the
Grants-in-Aid for Scientific Research No.\ 14079103 from the Ministry
of Education, Culture, Sports, Science and Technology, and No.\
15340088 from the Japan Society for the Promotion of Science.

\end{document}